\documentclass[pre,twocolumn,showpacs,showkeys,superscriptaddress]{revtex4-2}
\usepackage{color}
\usepackage{graphicx,hyperref,amsmath,soul}
\usepackage{mathtools,bm}
\usepackage{graphicx}
\usepackage{subfigure}
\usepackage{float}

\begin{document}

\preprint{APS/123-QED}

\title{Streamlined approach to mitigation of cascading failure in complex networks}

\author{Karan Singh}
\affiliation{School of Physics, Indian Institute of Science Education and Research Thiruvananthapuram, Thiruvananthapuram, 695551, Kerala, India.}
\author{V.K. Chandrasekar}
\affiliation{Department of Physics, Center for Nonlinear Science and Engineering, School of Electrical and Electronics Engineering, SASTRA Deemed
University, Thanjavur, 613401, Tamil Nadu, India.}

\author{D.V. Senthilkumar}%
\email{skumar@iisertvm.ac.in}
\affiliation{School of Physics, Indian Institute of Science Education and Research Thiruvananthapuram, Thiruvananthapuram, 695551, Kerala, India.}





\date{\today}

\begin{abstract}
Cascading failures represent a fundamental threat to the integrity of complex systems, often precipitating a comprehensive collapse across diverse infrastructures and financial networks. This research articulates a robust and pragmatic approach designed to attenuate the risk of such failures within complex networks, emphasizing the pivotal role of local network topology. The core of our strategy is an innovative algorithm that systematically identifies a subset of critical nodes within the network, a subset whose relative size is substantial in the context of the network's entirety. Enhancing this algorithm, we employ a graph coloring heuristic to precisely isolate nodes of paramount importance, thereby minimizing the subset size while maximizing strategic value. Securing these nodes significantly bolsters network resilience against cascading failures. The method proposed to identify critical nodes and experimental results show that the proposed technique outperforms other typical techniques in identifying critical nodes. We substantiate the superiority of our approach through comparative analyses with existing mitigation strategies and evaluate its performance across various network configurations and failure scenarios. Empirical validation is provided via the application of our method to real-world networks, confirming its potential as a strategic tool in enhancing network robustness.

\end{abstract}

\maketitle


\section*{Introduction}
Our everyday routines rely extensively on the operation of numerous natural and artificial networks, including neural and genetic regulatory networks, as well as communication systems, social networks, transportation networks, and electric power grids~\cite{watts1998collective,newman2010networks,latora2017complex}. However, these networks are susceptible to cascading failures, where the failure of a single node can lead to a domino effect, causing widespread breakdowns and disruptions in the entire network ~\cite{watts2002simple}. Understanding the resilience of these networks to random failures and targeted attacks is crucial for avoiding system failures with serious consequences~\cite{albert2000error}. Cascading phenomena can have either beneficial or adverse effects, depending on their practical usefulness. Why do certain works, such as books, movies, and albums, gain widespread popularity despite limited marketing efforts~\cite{gladwell2006tipping}, while similar endeavors fail to attract attention? Why do fluctuations occur in the stock market without any apparent significant news driving them~\cite{RePEc:aea:aecrev:v:85:y:1995:i:2:p:181-85}? How do grassroots social movements originate without centralized coordination or widespread public communication~\cite{396d471c-ddf9-39fd-9e67-b459b42aa9cd}? These occurrences exemplify what economists term as information cascades~\cite{RePEc:ucp:jpolec:v:100:y:1992:i:5:p:992-1026}, where individuals in a population exhibit herd-like behavior, basing their decisions on the actions of others rather than on their own information about the issue. Studies have also looked at how networks of adopters grow, particularly through social contagion ~\cite{rogers2005complex,pastor2015epidemic} processes like adopting opinions, behaviors, emotions ~\cite{vsuvakov2013online,tadic2013co,smiljanic2017associative,tadic2017agent} or innovations, which can happen quite rapidly under the influence of social pressure. Social influence is significant in this scenario, with individuals being influenced by the viewpoints of their peers~\cite{vsuvakov2013online,tadic2013co}. This impact is represented in threshold models~\cite{granovetter1978threshold,watts2002simple,easley2010networks}, that propose an individual becomes an adopter when the proportion of their already adopting neighbors reaches a critical level specific to their sensitivity.
Such phenomena underscore the complexities of human behavior and decision-making processes, particularly in contexts where social influence plays a significant role. One of the examples of the collapse of the top Online Social Networks(OSNs) is the Hungarian social networking site. The Hungarian social networking site iWiW~\cite{enwiki:707520580} was operational from 2004 to 2013. In late 2010, there was a noticeable increase in the number of users departing from iWiW, resulting in significant loss and ultimately leading to the collapse of the entire network. The main reason cited~\cite{torok2017cascading} for this downfall was that individuals left when most of their friends had already departed. 
Understanding the cascade effect is crucial for achieving various applications such as marketing, advertising, and spreading information. Identifying critical nodes where resources can be strategically allocated to effectively disseminate information across the network is key. However, these processes can also have negative implications in areas like finance, healthcare, and infrastructure by causing undesired or detrimental outcomes for the community. Therefore, it is important to mitigate the cascading failure in complex networks to minimize the potential negative consequences and ensure the stability and resilience of these systems.
\begin{figure*}[t]
    \includegraphics[width=0.9\textwidth]{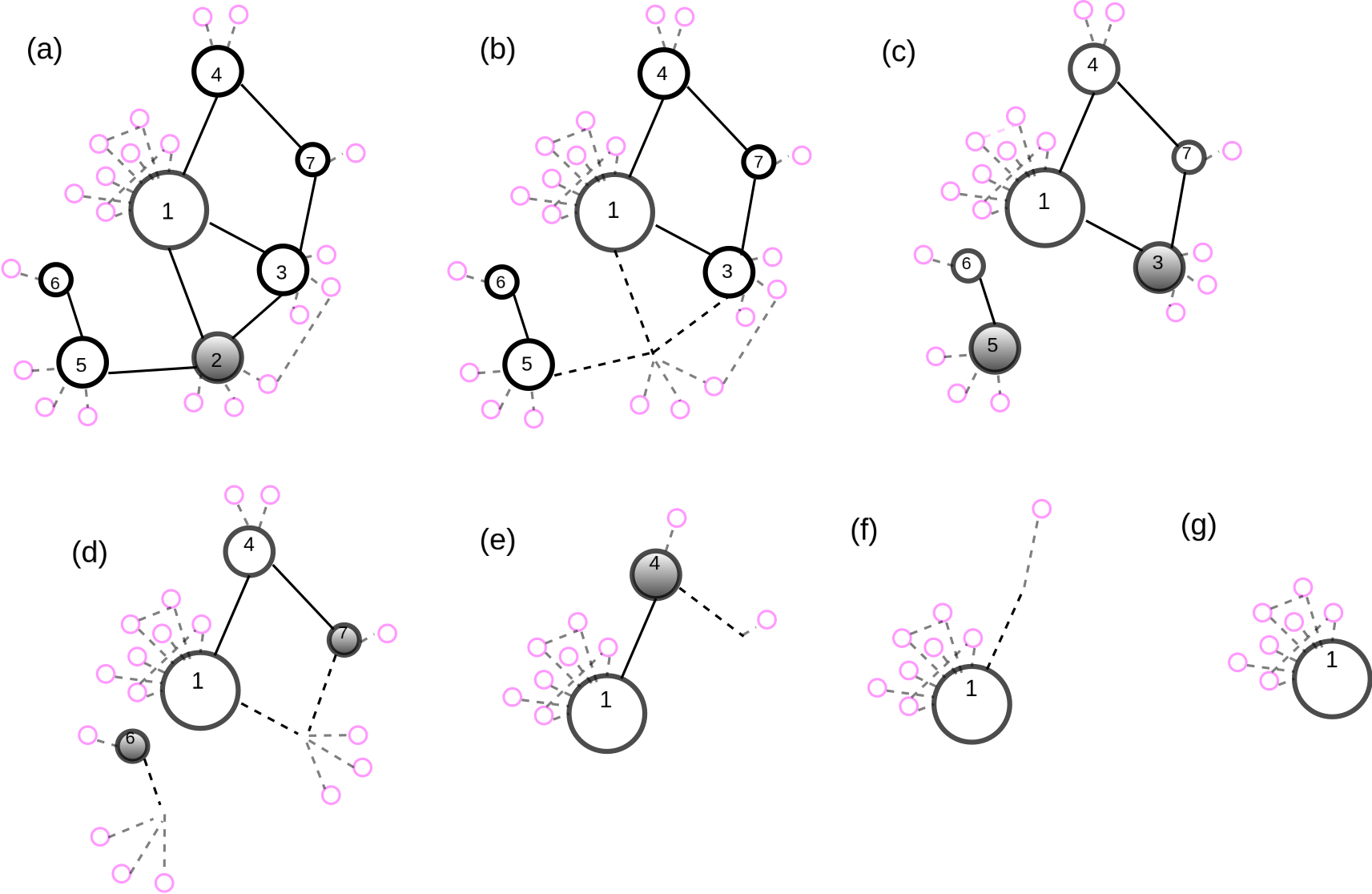}
    \caption{Illustration of the fractional cascade process. Node size corresponds to its degree, with only labelled nodes relevant to this description.
(a) The gradient (gray color) indicates the impacted node to be removed. (b) After removal, its edges are disconnected and eliminated. (c) Nodes are classified as high degree (1), low degree (6 and 7), and medium degree (3, 4, and 5). Low-degree nodes with few neighbors have minimal impact on the cascade. High-degree nodes remain unaffected by a single neighbor's failure, while medium-degree nodes
are susceptible to fractional failure and possess significant neighbors that will be impacted by their failure, hence intensifying the cascade.
 This subset of nodes exacerbates the cascading failure process. (d) to (h) depict the cascade progression, ultimately leaving only the
 high-degree node unaffected and surviving the failure.}
    \label{Fig1}
\end{figure*} 

In this work, we introduce a refined strategy for mitigating cascading failures, leveraging local environmental information of nodes, specifically focusing on the node's nearest neighbors~\cite{smolyak2020mitigation}. Our novel methodology involves the identification and enhancement of fragile nodes, the safeguarding of which ensures near-complete network survivability. To achieve this, we utilize graph coloring to ascertain the critical average degree, employing it as a threshold for refining the set of fragile nodes. Our investigation demonstrates that the proposed strategy is optimized in terms of both the fraction of fragile nodes and the probability of survival.

The effectiveness of our approach is first evaluated in theoretical models, namely Erd\H{o}s-R\'{e}nyi (ER) networks~\cite{erdds1959random} and Scale-free networks~\cite{albert2002statistical}, which emulate various real-world network structures. Subsequently, we assess its applicability in practical scenarios by demonstrating its efficacy in social networks, collaboration networks, and the US power-grid network. Our method not only adeptly mitigates cascading failures in simple theoretical models but also exhibits robustness in intricate real-life network contexts. We illustrate the versatility of our approach across a wide spectrum of network topologies, emphasizing its adaptability to more complex settings. Our findings underscore the efficacy of the proposed method in averting network collapse across diverse types of networks.
Understanding the dynamics of cascading failure in complex networks is crucial for developing effective mitigation strategies. As we continue to rely on and integrate complex networks into daily life, understanding and safeguarding against cascading failures becomes increasingly imperative.

\begin{figure*}[t]
    \includegraphics[width=0.9\textwidth]{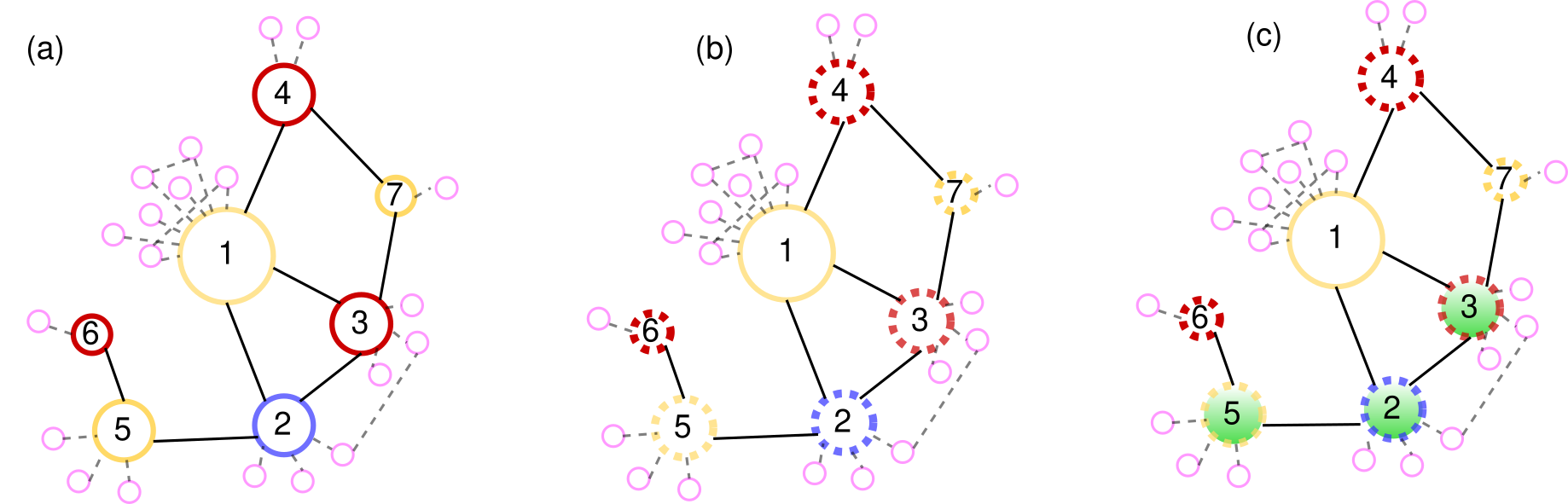}
    \caption{Visualization of the mitigation scheme.
(a) The graph is colored (ignoring the unlabelled nodes as they are not relevant). We are applying color to the periphery of the nodes
instead of fully coloring them. (b) Dotted-circumferenced nodes (2, 3, 4, 5, 6, and 7) are identified as fragile nodes under the fractional
 failure mechanism. (c) Among the fragile nodes, the green gradient nodes (2, 3 and 5)  with degrees exceeding the average lowest
degree (critical degree) of color groups are selected for protection, designated as immune nodes. When node 2 is impacted (refer to Fig. 1),
its inclusion in the protected set prevents its removal.  Consequently, its fragile neighbors (3 and 5) remain unaffected, and so on, ensuring the
 system's complete survival.}
    \label{Fig2}
\end{figure*}

\section*{The Models of Cascading Failure}
Many different failure models have been extensively studied and applied for specific purposes in simulating the propagation of failures. The k-core~\cite{chalupa1979bootstrap,centola2007complex}method and the fractional threshold are two common approaches heavily used to represent real-life scenarios. In the k-core method, the threshold is determined by the absolute number of active neighbors, making it particularly applicable in epidemiological settings where an individual's actual number of contacts is crucial. On the other hand, when the fraction of active neighbors holds more significance than their absolute number, it becomes known as the fractional threshold~\cite{huang2013cascading,watts2002simple,gai2010contagion,feng2015competing} model. This model finds application in various fields such as opinion formation, social dynamics, infrastructures, economics, and finance.
In the context of opinion formation, an individual adopts a certain behavior if a fraction $m$ of its friends $k$ adopt it. In this case, the threshold is $m/k$ and not $m$.  This results in individuals with many friends needing more of their neighbors to become adopters to change their state. This process is illustrated in Fig. \ref{Fig1}. 

Watts~\cite{watts2002simple} presented a simple model on global cascades, implementing the fractional threshold model for infinitesimally small impact. Later, Gleeson and Cahalane~\cite{gleeson2007seed} modified it for finite seed size and studied its effects on the cascades.

We used the Gleeson et al.~\cite{gleeson2007seed} framework for analytic calculations of cascading failure processes. Each node, having degree $k$ with the degree distribution $p_k$ such that $\sum p_k = 1$, is an agent that belongs to the undirected network and can be in a binary state, called \textit{active} or \textit{inactive}. All agents are assigned a frozen random threshold $r$ chosen from the distribution with $F(r)$ denoting the probability that an agent has a threshold less than $r$. The cascade is initiated by seeding the network by activating a randomly chosen fraction of nodes $\rho_0$ out of total $N$ nodes. Nodes update their state, and the average final fraction $\rho$ of active nodes is given by~\cite{gleeson2007seed} 
\begin{equation}
    \rho = \rho_0 + (1-\rho_0)\sum_{k=1}^{\infty}p_k \sum_{m=0}{k}\begin{pmatrix} k \\ m \\ \end{pmatrix}q_{\infty}^m (1-q_{\infty})^{k-m}F\left(\frac{m}{k}\right)
\end{equation}
where $q_\infty$ is the steady state or fixed point of the recursion relation 
\begin{equation}
    q_{n+1}=\rho_0+(1-\rho_0)G(q_n) \hspace{1cm}    n =0,1,2,\dots
\end{equation}
and the generating function $G$ is defined as 
\begin{equation}
    G(q_n)=\sum_{k=1}^{\infty}\frac{k}{z}p_k\sum_{m=0}^{k-1}\begin{pmatrix} k-1 \\ m \\ \end{pmatrix}q_n^m (1-q_n)^{k-1-m}F\left(\frac{m}{k}\right)
\end{equation}
Here $z$ is the network's average degree $z=\sum kp_k$.

\begin{figure}[h]
    \includegraphics[width=0.4\textwidth]{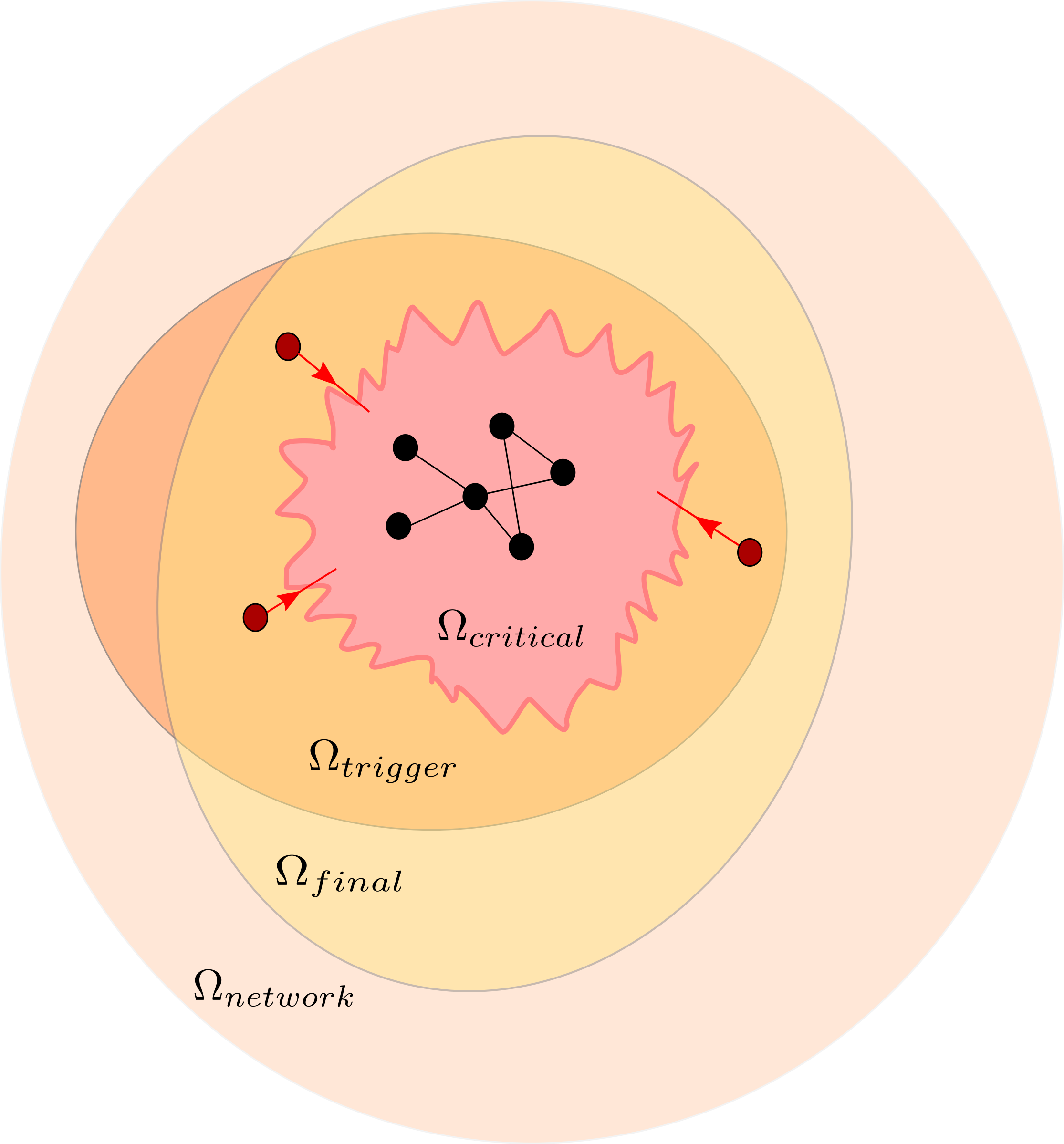}
    \caption{Critical Mass: In every network exists a critical mass $\Omega_{critical}$ where numerous vulnerable nodes are interconnected. These nodes only require one of their friends to activate them so they can also switch on. Just beyond the critical mass component lies the triggering component $\Omega_{trigger}$, where certain nodes are connected to critical nodes. When these connecting nodes (depicted as red nodes in Fig. 3) become active, they trigger someone within the critical component and initiate a network collapse. $\Omega_{final}$ is the fraction of failed nodes after the cascade is settled, and $\Omega_{network}$ is the entire network.}
    \label{Fig3}
\end{figure} 

\section*{Methods}
\subsection*{Mitigation strategy}

As discussed in a recent study~\cite{smolyak2020mitigation}, our focus will be on identifying critical nodes that play a key role in propagating cascading effects instead of measuring a node's importance based on the damage it causes when removed. As shown~\cite{smolyak2020mitigation}, the node's importance in terms of the damage it causes doesn't help from the mitigation perspective beacause at a critical threshold, even an exceedingly small impact~\cite{smolyak2020mitigation} (as minor as removing a single node) could initiate (see Fig. \ref{Fig3}) a cascade effect leading to network collapse.

We adhere to the concept~\cite{smolyak2020mitigation} of a fragile node within a failure mechanism, $M$, wherein the failure of the node occurs upon the removal of a minimal significant number, $K$, of its links. The fractional failure under the removal of a single edge takes the form $(k-1)/k$, and the node will be considered to fail if $(k-1)/k < \theta$.
We will now present the algorithm to select the potent nodes based on the basic percolation argument: to facilitate the cascade, the node must be fragile under the fragility condition (mentioned above) and possess at least some fragile neighbours. 
\\ 




\begin{tabular}{|p{8cm}|}
    \hline
    \textbf{Algorithm: Outputs nodes to be protected} \\
    \hline
    \begin{enumerate}
        \item Scan each node
        \item Check if it is fragile as per the above condition
        \item If yes, check if it has at least two fragile neighbours
        \item Color the graph and calculate the average degree of each color subgroup
        \item Protect only those fragile nodes (fulfilling 2\textsuperscript{nd} and 3\textsuperscript{rd} condition) with degrees greater than the lowest average degree color group.
    \end{enumerate} \\
    \hline
    \end{tabular}

\vspace{1cm}
The first two steps are basics and are to determine the fragile nodes, with the rationale being that, as per the fragility defined above, i.e., a node fails under the removal of the single edge, but if the node remains intact, it is unlikely to be affected at the initial cascade stage as it is not fragile enough as per the definition. Additionally, the cascade doesn't aggravate if the node's fragile neighbours are fewer than two in number. The last two steps are to reduce/refine the number of fragile nodes. It's found (see the Results section) that there is no need to protect all the fragile nodes, but protecting a fraction(which we achieve after the application of our last step) would ensure the full survivability of the network. We refer to this reduced/refined set as immune nodes (the set of nodes to be protected). We use the graph coloring~\cite{formanowicz2012survey} technique to reduce the number of fragile nodes (more discussion in the Result section). Upon examining the algorithm, it is observed that nodes with high degrees are not suitable for inclusion in the immune set because they remain resilient even under very high thresholds. On the other hand, low-degree nodes become fragile relatively quickly, but they do not effectively transmit the shock and, therefore, cannot be considered for the immune set. The ideal candidates for immune nodes lie between high and low degrees – these medium-degree nodes are sufficiently weak to be fragile and well-connected enough to propagate the shock significantly. Selecting this group of nodes offers an effective mitigation strategy due to their limited number.

\begin{figure}[h]
    \includegraphics[width=0.5\textwidth]{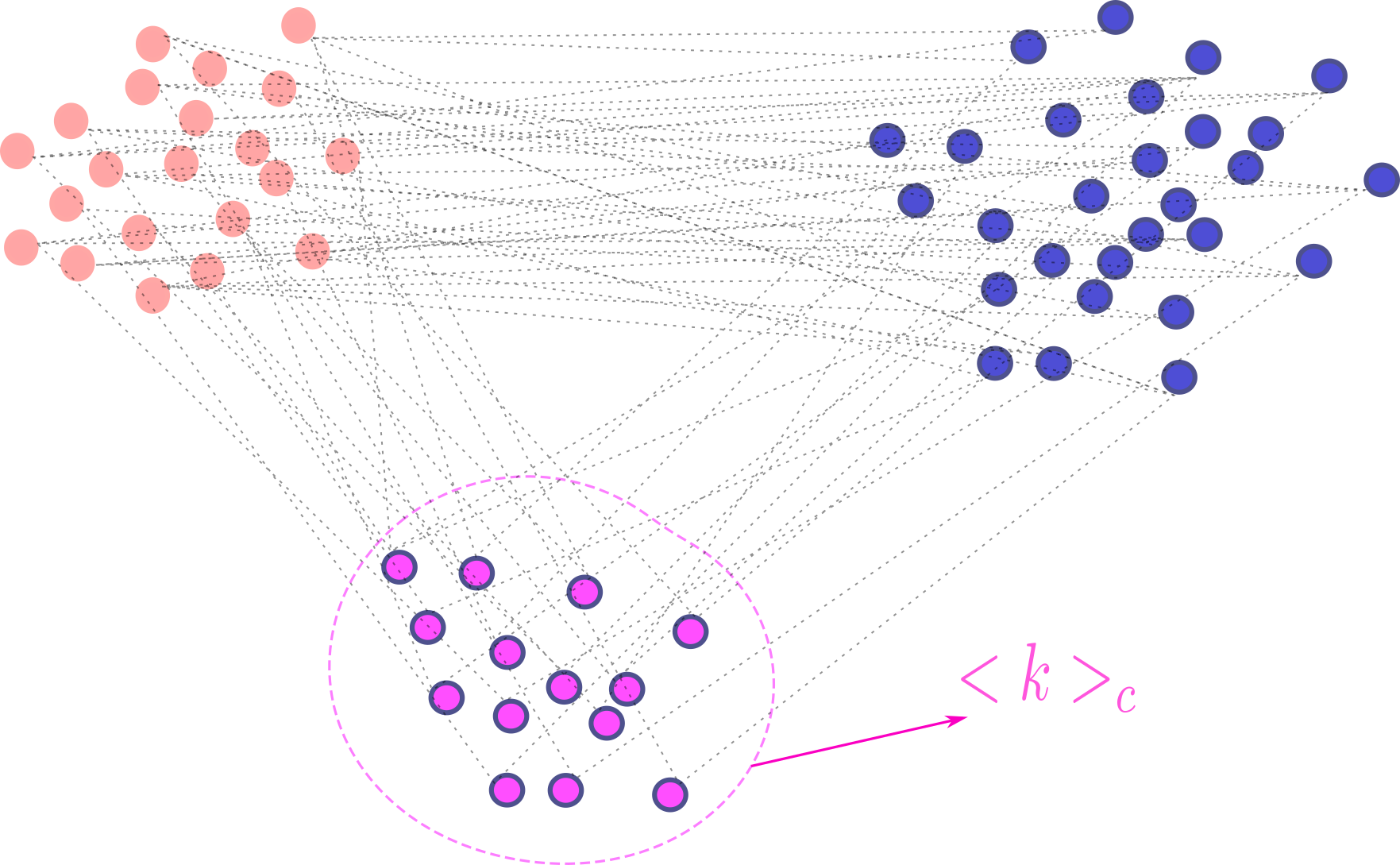}
    \caption{Critical degree: Coloring the graph transforms it into $k$-partite graph. Its vertices are partitioned into $k$ different independent sets (we refer to them as color groups), where $k$ is the chromatic number, for example, $k=2$ is the bipartite graph. $<k>_c$ is the lowest average degree among the color groups. Here, we have taken a $k=3$ example for the demonstration purpose .}
    \label{Fig4}
\end{figure} 

\section*{Results}
\subsection*{Application on standard networks}
The primary objective of the algorithm is to pinpoint the critical nodes in terms of failure propagation. The selection of nodes relies on understanding their local surroundings, making it a highly effective mitigation strategy with minimal and readily accessible information. Using this group of vulnerable nodes, we apply graph coloring~\cite{formanowicz2012survey} techniques to minimize this number and create a more cost-effective strategy. We refer to this subset as immune nodes (as the nodes to be protected; see green nodes in Fig. \ref{Fig2}).
Coloring the graph and dividing it into subgroups (colour groups), allows us to determine the connectivity of a fragile node with the other colour groups based on the average degree of that particular colour group. For example, suppose the average degree of a certain colour group is high. In that case, each node is highly connected with other colour groups but not at all connected among the same colour group. If this node is determined to be fragile, upon removal of this node, it will affect the other nodes in different color groups, thereby endangering their fractional threshold. Therefore, by choosing a threshold above which the node designated for protection possesses the lowest average degree among the color groups, we effectively and automatically safeguard the fragile nodes characterized by high interconnectivity between different color groups. \\
We first apply our algorithm to the standard network models, Erdos-Renyi and Scale-free networks, to demonstrate its effectiveness.
Figures 5 (a) and (c) show the survivability of the system against the fractional threshold. The value on the y-axis represents the probability of the system survival, and value on the x-axis denotes the fractional threshold, that means the fraction of functional neighbors required for a node in question for its survival. The blue line represents full protection (1.0), which means protecting all the nodes (immune nodes) selected by our algorithm. The dashed pink line, corresponding to the full protection case, indicates the fraction of protected nodes compared to the entire network. The orange (0.7) and green (0.4) lines indicate partial protection - that is, protecting the immune nodes 70\% or 40\% of the times respectively. Finally, the red line shows no protection; it depicts how the system evolves without any form of safeguarding in place. The system always survives until the fractional threshold of 0.86, regardless of the protection. Therefore, for the ER case, 0.86 is considered the critical threshold. As for scale-free networks, it's close to 0.835. Near the critical threshold, it just requires a small fraction of nodes to be protected for the complete survival of the network, as shown by the mitigation strategy (blue line) and basic cascading process (yellow line) in Figures 5 (a) and (c).

Figures 5 (b) and (d), for ER and Scalefree networks respectively,  show the system's survivability for all protection probabilities plotted for four different fractional thresholds depicted by four colours. The blue lines correspond to a fractional threshold that is lower than their respective critical threshold, ensuring the system always survives. As the threshold increases, the probability of survival decreases for the orange, green, and red lines. Increasing the protection probability enhances the survivability of the system. To guarantee that the system always survives, it is necessary to protect the immune nodes with a protection probability of 1.0 (as depicted in the Fig. 5 (b) and (d)).

We have demonstrated the efficacy of our algorithm on ER and Scalefree networks. It identifies the fragile nodes and substantially reduces their number, obtaining what we call immune nodes. These immune nodes are designated for the protection and ensuring the full survival of the system. To show that the reduce/refine step of our algorithm is optimised in terms of both survival probability and the protected node size, we scale the critical degree $<k>_c$ (See Fig. \ref{Fig4}) by lowering and raising its value to test whether this automatic selection of the critical degree threshold is optimised or not. In Figure \ref{Fig6}, the critical degree threshold for (a) ER and (b) Scale-free networks is scaled. The green solid line in Fig.\ref{Fig6}(a) represents the survival probability for a fully protected network with the original critical threshold degree, while the corresponding green dashed line denotes the protected set. The blue and yellow lines indicate thresholds scaled to 95\% and 97\% of the original threshold degree, respectively, allowing more nodes (as shown by corresponding dashed lines) to be included in the protected set, resulting in improved survival probability. However, raising the original threshold to 102\% or above (as seen by solid red or magenta lines) results in poor survivability due to the reduction of nodes in the protected set (shown by corresponding dashed lines). 

\begin{figure*}[t]
    \includegraphics[width=0.7\textwidth]{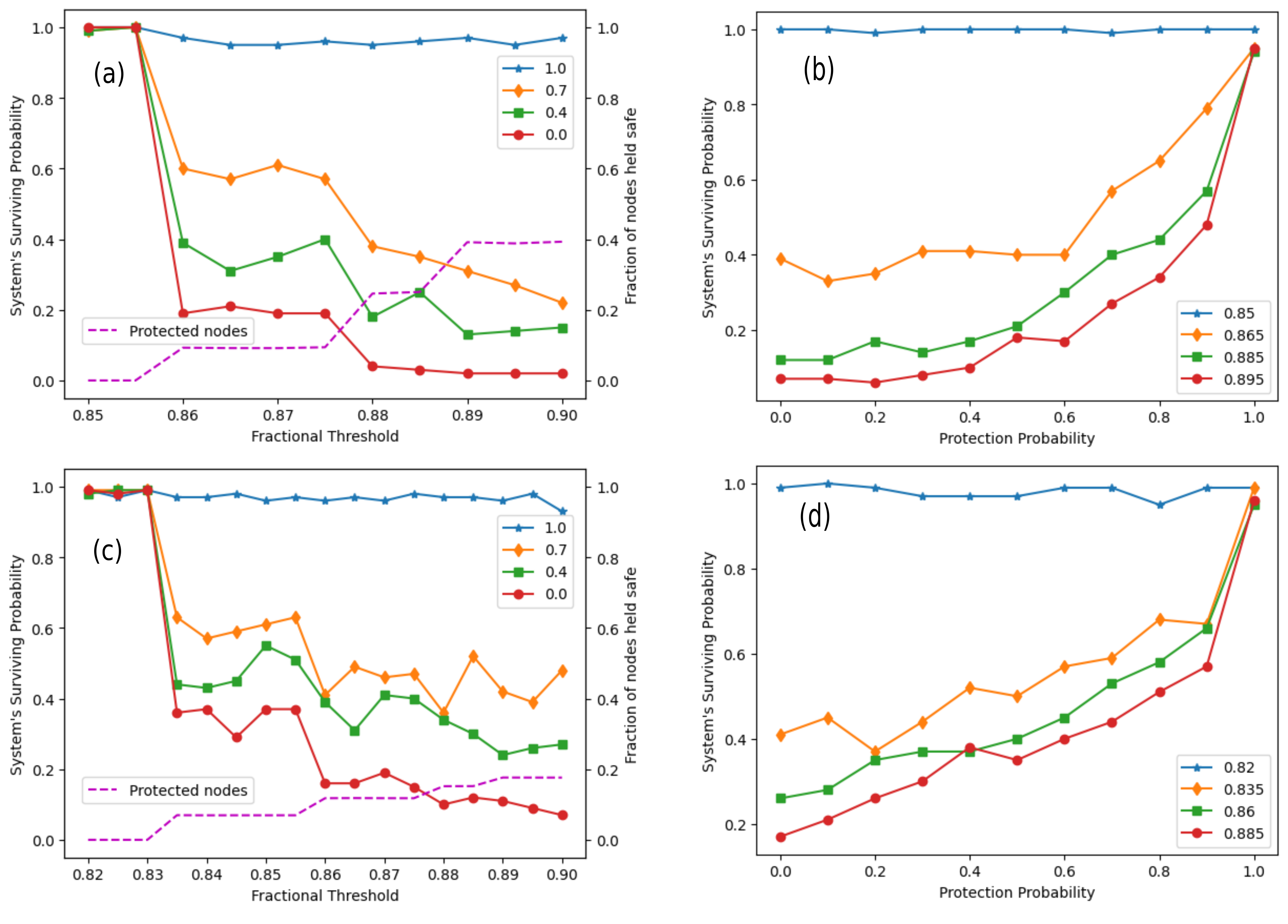}
    \caption{Mitigation of cascading failure on standard ER and Scalefree networks. (a) The plot shows the surviving probability vs Fractional threshold for ER network ($N=20000$ and $<k>=8$) with four protection probabilities - Blue, orange, green and red representing probabilities 1.0, 0.7, 0.4, and 0.0, respectively. The blue curve represents a fully protected network, while the red line indicates no protection. The orange and green curves represent partial protection. Additionally, the pink dashed line denotes the size of immune nodes (only for fully protected case) as a fraction of the entire network (right y-axis). (b) Surviving probability is plotted against all protection probabilities for an ER network using four different fractional thresholds. (c,d) Similar configurations are plotted for Scalefree network.}
    \label{Fig5}
\end{figure*} 

\begin{figure*}[t]
    \includegraphics[width=0.7\textwidth]{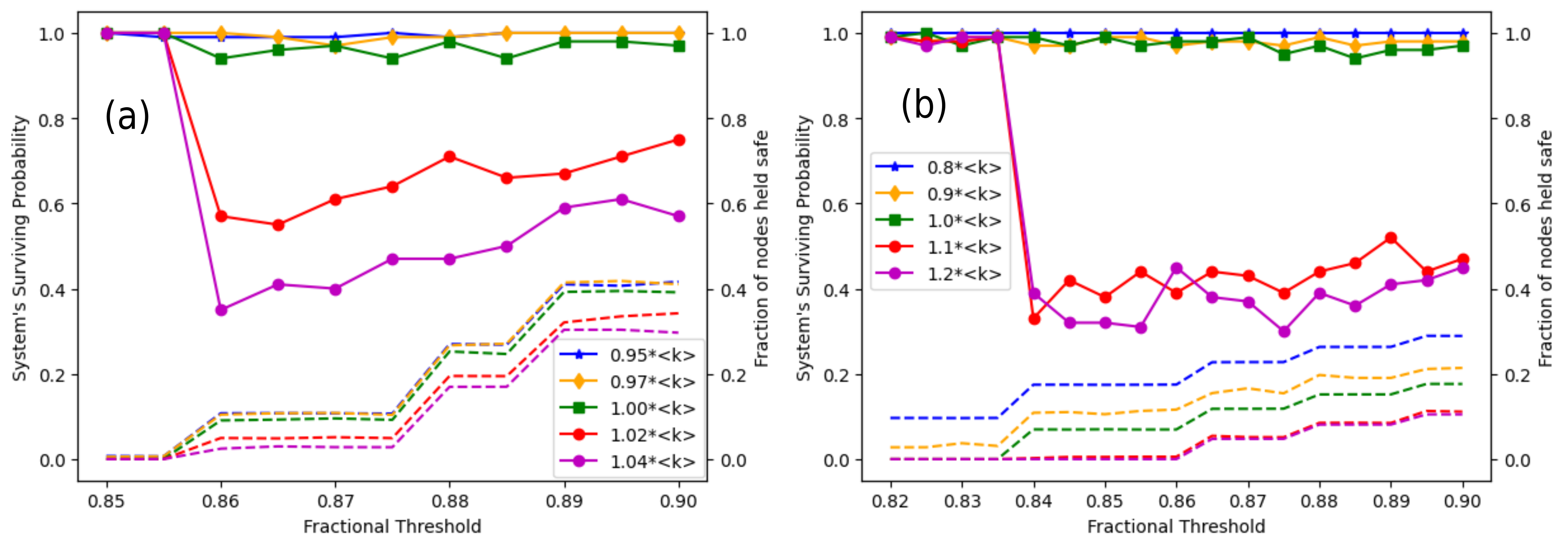}
    \caption{Scaling the critical degree threshold for (a) ER and (b) scalefree networks for full protection case. The solid green lines in both figures have the original critical threshold (i.e., the lowest average degree of the color groups). The dashed lines represent the respective protected set as the fraction of the entire network size (right axis). The solid yellow and blue lines represent the lower critical threshold, allowing more nodes (see dashed yellow and blue lines) to be protected and resulting in a higher survival probability. The solid red and magenta lines represent an increase in the critical threshold, allowing fewer nodes (see dashed red and magenta lines) to be protected and thus reducing the survival probability.
    }
    \label{Fig6}
\end{figure*} 

Similar scaling was tested for scale-free networks, revealing that ER is more sensitive than Scale-free networks in terms of scaling factor. So, this demonstrates that our refining/reducing step in the algorithm effectively establishes the threshold degree to which fragile nodes with a degree greater than that should be protected to enhance the system's surviving probability.
 \begin{figure*}[t]
    \includegraphics[width=0.6\textwidth]{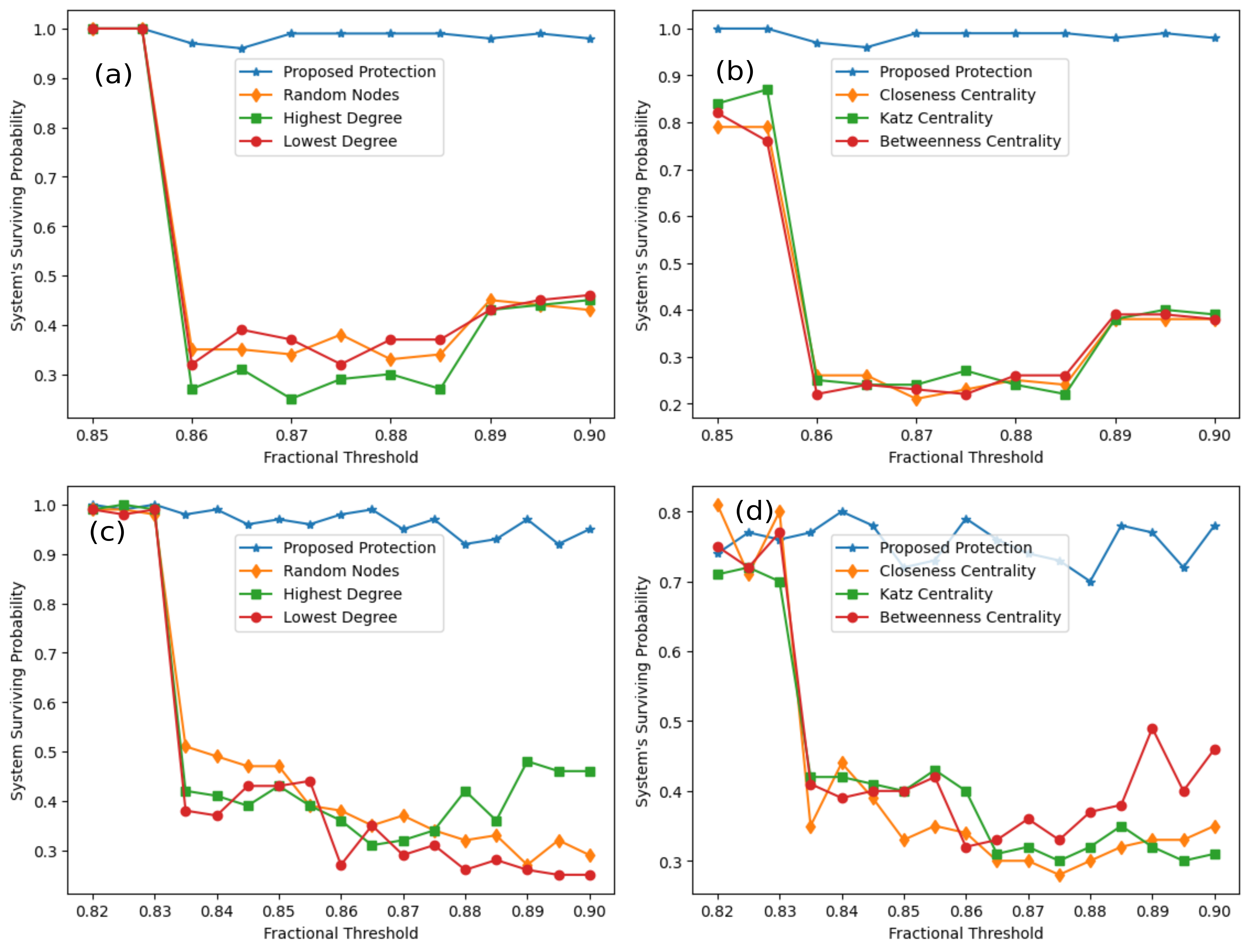}
    \caption{Comparison with the standard mitigation techniques. (a,b) ER (c,d) Scalefree
    }
    \label{Fig7}
\end{figure*} 
\subsection*{Comparison with standard mitigation techniques.}
Fig. \ref{Fig7} illustrates the effectiveness of our algorithm in comparison to standard intuitive and centrality-based mitigation strategies. Initially, we compare our proposed technique with intuitive algorithms. In Fig. \ref{Fig7} (a), we compare our proposed protection for ER networks with simple intuitive strategies using the same set size of protected nodes for a fair comparison.

We start by randomly selecting and protecting nodes to verify the effectiveness of the proposed strategy compared to random selection,the point here is to verify if the proposed strategy is actually effective or is it just another random selection. As shown clearly, it's much more than the random selection as it outperforms it. The next obvious candidate is to protect the high-degree nodes; the reason behind this is that since it is highly connected with the other nodes, keeping them safe would protect their neigbors. That, too clearly is not the reasonable set to protect, as clearly shown in the figure.
Protecting the lowest degree nodes, the rationale behind it is that saving the most vulnerable nodes in the network ensures the better survival of the network.

Fig. \ref{Fig7} (b) compares our proposed protection for the ER networks with the centrality-based strategies. A better insightful class for selecting nodes would be different centrality measures: closeness centrality, Katz centrality, and betweenness centrality. Closeness centrality measures how close a node is to all other nodes in a network, indicating nodes that can quickly interact with others. Katz centrality quantifies the influence of a node in a network by considering the number of immediate neighbors as well as the neighbors' influence recursively throughout the network, and Betweenness centrality identifies nodes that act as bridges between different parts of a network, measuring the fraction of shortest paths that pass through a node, indicating its importance in facilitating communication or information flow. Once again, our proposed strategy stands out, providing enhanced survival chances compared to these centrality measures techniques. We obtained the same results for the Scalefree network as shown in the Fig.\ref{Fig7} (c,d).

\subsection*{Application to the Real-World networks}
\subsubsection*{1. LastFM Asia Social Network}

\begin{figure*}[t]
    \includegraphics[width=0.6\textwidth]{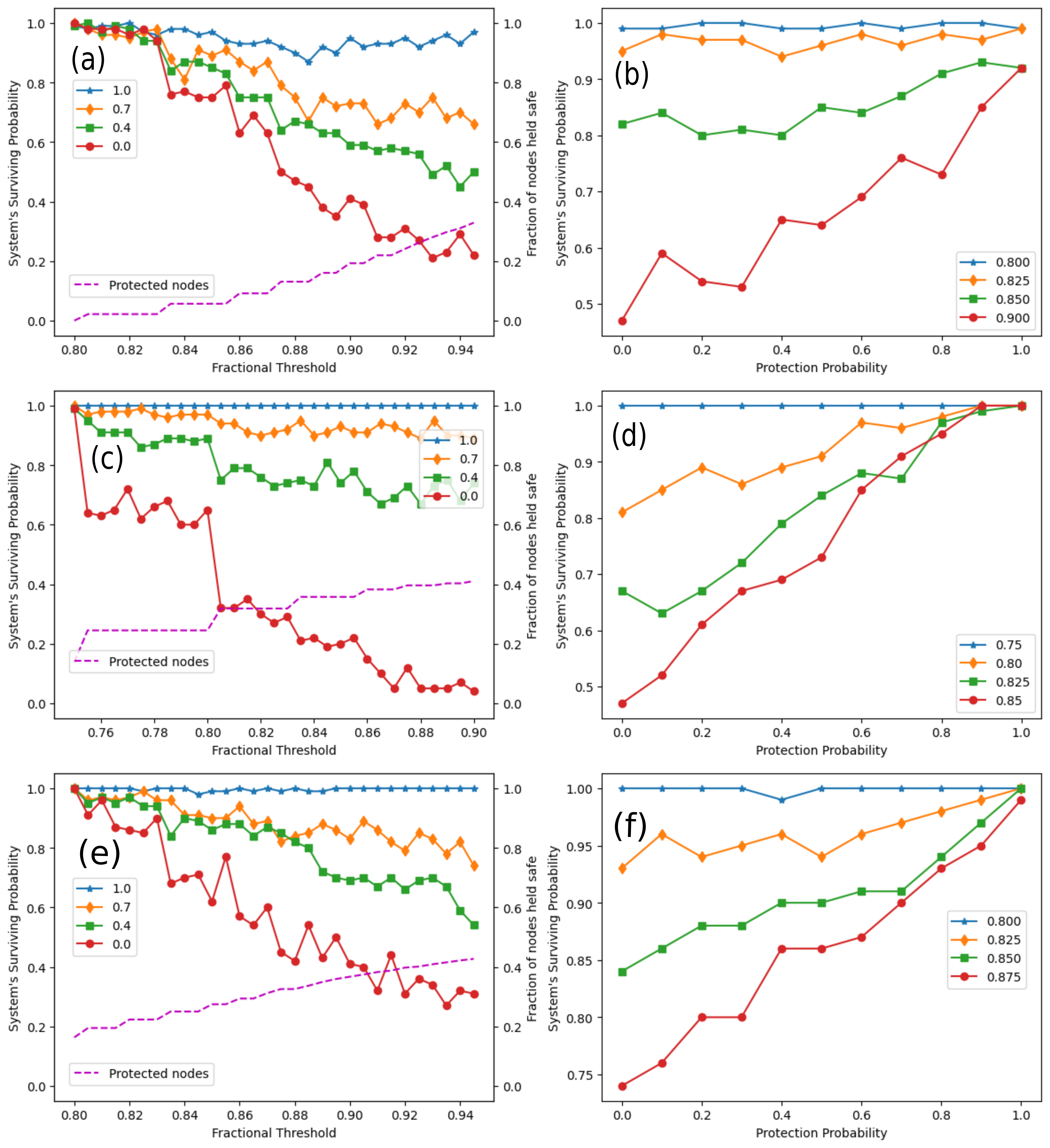}
    \caption{Application to the Real-World Networks. (a-b) LastFM Social Network, where the nodes are the users and mutual follower relationships are the links. (c-d) US Power Grid network where nodes within the network correspond to essential components such as generators, transformers, or substations, while edges symbolize the interconnected power supply lines.(e-f) Condensed matter physics collaboration network, authors are represented as nodes. If authors have collaborated, there exists a link between them. When more than two authors (denoted as $k$) are involved, a connected subgraph is created among these $k$ authors (a) The plot shows the surviving probability vs Fractional threshold for the LastFM Social Network ($N=7626$ and $<k>=7.29$) with four protection probabilities - Blue, orange, green and red representing probabilities 1.0, 0.7, 0.4, and 0.0, respectively. The blue curve represents a fully protected network, while the red line indicates no protection. The orange and green curves represent partial protection. Additionally, the pink dashed line denotes the size of immune nodes (only for fully protected case) as a fraction of the entire network (right y-axis). (b) Surviving probability is plotted against all protection probabilities for the LastFM Social Network using four different fractional thresholds. (c,d) The same analysis is carried out for the US Grid network ($N=4941$ and $<k> = 2.67$). 
    }
    \label{Fig8}
\end{figure*}

We proceed to apply our mitigation framework to assess its efficacy in a real-world scenario. First, we examine its application within the LastFM Asia Social Network—a collection of LastFM users sourced from the public API in March 2020. In this network, nodes represent LastFM users hailing from Asian countries, while edges denote mutual follower relationships between them. The number of user recorded is $N=7626$, and the average degree $<k>=7.29$. Fig. \ref{Fig8} (a,b) shows the result of the application of our mitigation strategy on the real-world data of the social network site. (a) System's survivability is plotted against the fractional threshold for four different protection probabilities, and in (b), the same is plotted for all protection probabilities. 

The fractional threshold model proves pertinent in delineating the dynamics of collapse within Online Social Networks (OSNs). Users tend to disengage from these platforms when a substantial portion of their social connections has already withdrawn from active participation. The waning interest of an individual in a social networking site may stem from its failure to furnish compelling features, content, or interactive activities. Consequently, users may curtail or cease their habitual sharing of daily experiences on the platform, or altogether abandon its usage. This decline in activity can precipitate a ripple effect, causing the user's network connections to lose interest due to the absence of their friend's contributions or the inability to engage in real-time interactions. If a user observes that a significant proportion, denoted by $m$ out of $k$, of their network connections have exited the platform, it may prompt their own departure. This phenomenon can propagate through the network, culminating in a cascade failure. Notably, the Hungarian social networking site iWiW~\cite{enwiki:707520580} fell victim to such a cascade failure. Operating from 2004 to 2013, iWiW witnessed a surge in user departures towards the end of 2010, resulting in substantial attrition and eventual network collapse. Research~\cite{torok2017cascading} attributes this decline to individuals departing once their social circles had largely vacated the platform. In response, major OSNs such as Facebook and Instagram have strategically integrated engaging features like platform games, activity sharing, story updates, and enhanced chat functionalities to mitigate similar attrition trends and bolster user engagement.


\subsubsection*{2. US Power Grid Network}
The power grid network is mostly based on the redistribution of load~\cite{zhang2015vulnerability}. Here, we model it very basics by fractional threshold model. That is, the grid fails if more than $m$ out of its $k$ connecting grids fails. We are utilizing the US grid network with nodes $N = 4941$ and average connectivity $<k> =2.67$. Fig. \ref{Fig8} (c,d) shows the result of the application of our mitigation strategy on the real-world data of the US Grid network. (a) System's survivability is plotted against the fractional threshold for four different protection probabilities, and in (b), the same is plotted for all protection probabilities. In this context, nodes within the network correspond to essential components such as generators, transformers, or substations, while edges symbolize the interconnected power supply lines. In the event of a transformer or substation malfunction, the repercussions extend to the connected transformers, as they assume the burden of redistributed loads. Employing a fractional threshold to gauge the vulnerability of a transformer based on the proportion of its neighboring components could be a pragmatic approach. By leveraging such a strategy, the algorithm can proactively identify susceptible transformers for preemptive upgrades, mitigating the risk of future load fluctuations or overload failures. 

\subsubsection*{3. Collaboration Network}

Collaboration networks are primarily Scalefree in nature. The fractional threshold model could be highly useful for analyzing the failure of an institution, industry, or researcher. Here, we utilize the condensed matter physics collaboration data with nodes $N = 23133$ and average degree $<k> = 8.08$ that is derived from the e-print arXiv and encompasses scientific collaborations among authors who have submitted papers to the Condensed Matter category. In this network, if author $i$ has co-authored a paper with author $j$, an undirected edge is included between $i$ and $j$. Moreover, when a paper is co-authored by $k$ authors, this results in a fully connected subgraph comprising $k$ nodes.

The rationale is that an individual researcher needs at least some of their collaborators to be present and participate in the publication. If this does not happen, it could result in a decrease in the rate or quality of publications, or publication may even stop altogether. Collaboration failure can occur due to various reasons such as severed ties, illness or death of a collaborator, non-compliance with mutual agreements, issues related to credit sharing, and more.
As depicted in Fig.\ref{Fig8} (e), the red line shows that randomly removing a few researchers (two in this case, in order to start the cascade) from the network affects their immediate neighbors and leads to a cascade of failures until the network collapses. On the other hand, the blue line demonstrates that protecting a fraction of nodes (pink dotted line), such as collaborators, ensures the survival of the network. The orange and green lines represent probabilistic protection of critical nodes by 70\% and 40\%, respectively, significantly increasing the probability of survival for partially protected networks. And in Fig. \ref{Fig8} (b), the system's surviving probability is plotted for all protection probabilities. 

\section*{Discussion and Conclusion}

In this work, we introduce an algorithm designed to efficiently mitigate cascading failures in complex networks. This mitigation is achieved by safeguarding critical nodes that are most effective at propagating and intensifying failure. Differently from the existing work on the mitigation of cascading failures in
complex networks~\cite{smolyak2020mitigation}, we have exploited the graph coloring concept to identify the critical degree threshold, designating nodes with a degree greater than it for protection. This approach optimally refines the fragile nodes without compromising the system's survival probability. Our findings demonstrate that effective node selection can be accomplished with only minimal understanding of the local neighborhoods of the nodes and the underlying failure mechanisms. This strategy significantly enhances the likelihood of network resilience, even without precise knowledge of the initial source of impact. We have evaluated our method across various network configurations and failure scenarios, yielding robust mitigation strategies. Furthermore, we have successfully implemented this approach in real-world network settings, including social networks, the US grid network, and collaboration networks. In each case, our algorithm has effectively mitigated cascading failures. However, it leaves some open questions for further research.

\bibliography{main}

\clearpage
\setcounter{figure}{0}
\renewcommand{\figurename}{Fig.}
\renewcommand{\thefigure}{S\arabic{figure}}

\setcounter{equation}{0}
\renewcommand{\theequation}{S\arabic{equation}}
\onecolumngrid

\begin{center}
    {\LARGE{Supplementary Information} \\ \Large{Streamlined approach to mitigation of cascading failure in complex networks}}
\end{center}

\section{Analytical calculation of cascading failure processes}

We apply the Gleeson et al.~\cite{gleeson2007seed} framework for analytic calculations of cascading failure processes. Each node, having degree $k$ with the degree distribution $p_k$ such that $\sum p_k = 1$, is an agent that belongs to the undirected network and can be in a binary state, called \textit{active} or \textit{inactive}. All agents are assigned a frozen random threshold $r$ (uniform threshold for our case) chosen from the distribution with $F(r)$ denoting the probability that an agent has a threshold less than $r$. The cascade is initiated by seeding the network by activating a randomly chosen fraction of nodes $\rho_0$ out of total $N$ nodes. Nodes update their state, and the average final fraction $\rho$ of active nodes is given by~\cite{gleeson2007seed}

\begin{equation}
    \rho = \rho_0 + (1-\rho_0)\sum_{k=1}^{\infty}p_k \sum_{m=0}{k}\begin{pmatrix} k \\ m \\ \end{pmatrix}q_{\infty}^m (1-q_{\infty})^{k-m}F\left(\frac{m}{k}\right)
\label{Eq1}
\end{equation}
where $q_\infty$ is the steady state or fixed point of the recursion relation 
\begin{equation}
    q_{n+1}=\rho_0+(1-\rho_0)G(q_n) \hspace{1cm}    n =0,1,2,\dots
\end{equation}
and the generating function $G$ is defined as 
\begin{equation}
    G(q_n)=\sum_{k=1}^{\infty}\frac{k}{z}p_k\sum_{m=0}^{k-1}\begin{pmatrix} k-1 \\ m \\ \end{pmatrix}q_n^m (1-q_n)^{k-1-m}F\left(\frac{m}{k}\right)
\end{equation}
Here $z$ is the network's average degree calculated as $z=\sum kp_k$.
The improved cascade condition developed by Gleeson et al.~\cite{gleeson2007seed} for a finite initial impact,
\begin{equation}
   (C_1 -1)^2 - 4C_0C_2 + 2\rho_0(C_1-C_1^{2}-2C_2+4C_0C_2) < 0
\label{Eq2}
\end{equation}
where $C_l$'s are the coefficients of power series $G(q) = \sum_{l=0}^{\infty}C_lq^l$.

\begin{figure}[h]
    \centering
    \includegraphics[width=0.6\textwidth]{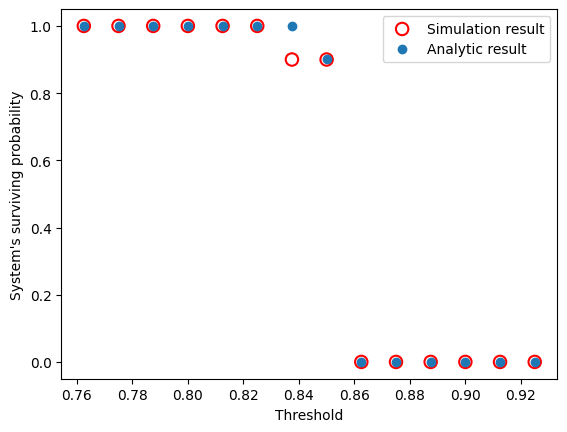}
    \caption{Demonstration of the fractional cascade process on a ER network with $N=20000$ nodes and average degree $<k>=8$.}
    \label{SI_1}
\end{figure}
\begin{figure}[h]
    \centering
    \includegraphics[width=0.6\textwidth]{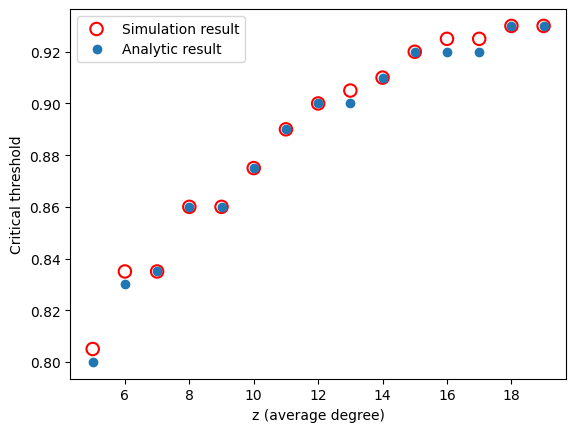}
    \caption{Analytical and simulation results of the critical fractional threshold compared to the initial average degree of ER network. }
     \label{SI_2}
\end{figure}

Fig. \ref{SI_1} illustrates the process of cascading failure and simulation results are in good fit with the analytical result (Eq. \ref{Eq1}) of an ER network. A collapse of the network occurs at the critical threshold around 0.86. In Fig. \ref{SI_2}, the critical threshold for various initial average degrees is plotted, showing good alignment with the analytical result (\ref{Eq2}). As the average degree of the network increases, so does its critical threshold; consequently, it will collapse at a higher threshold level.

\noindent

\begin{figure}[h]
    \centering
    \includegraphics[width=0.8\textwidth]{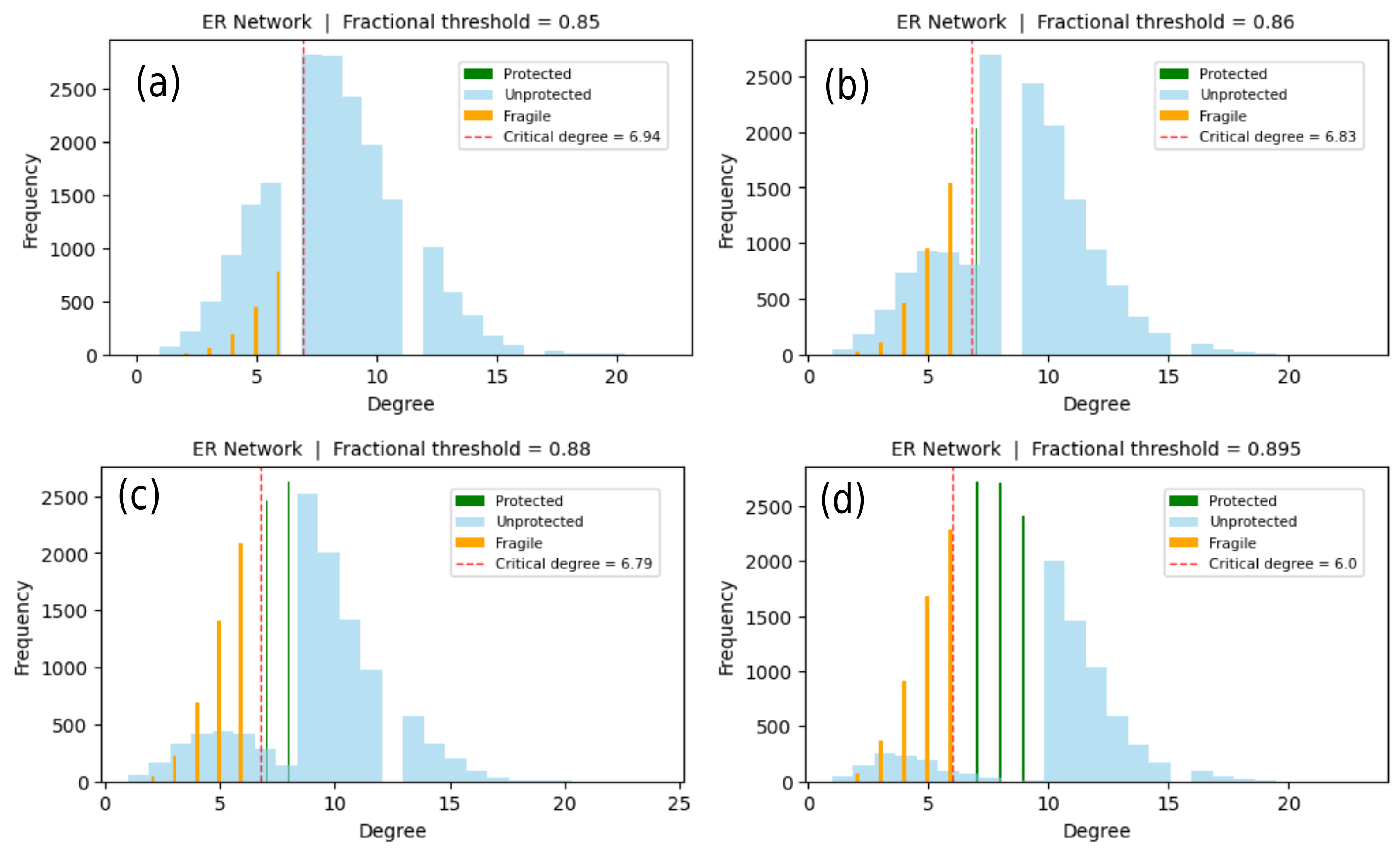}
        \caption{Frequency distribution of node types are observed at four different fractional thresholds for the ER network consisting of $N =20000$ nodes and average degree $<k>=8$. The sky blue bars represent the unprotected or non-fragile nodes, while the yellow bars indicate fragile nodes. A red dashed line denotes the critical degree above which the fragile nodes should be protected. (a) At a fractional threshold of 0.85, no protection is required as all fragile nodes are below the critical threshold (dashed-red line). (b-d) From thresholds 0.86 to 0.89, an increasing number of fragile nodes lie above the critical threshold and therefore need to be protected (green bars). }
    \label{SI_3}
\end{figure}

\begin{figure}[h]
    \centering
    \includegraphics[width=0.8\textwidth]{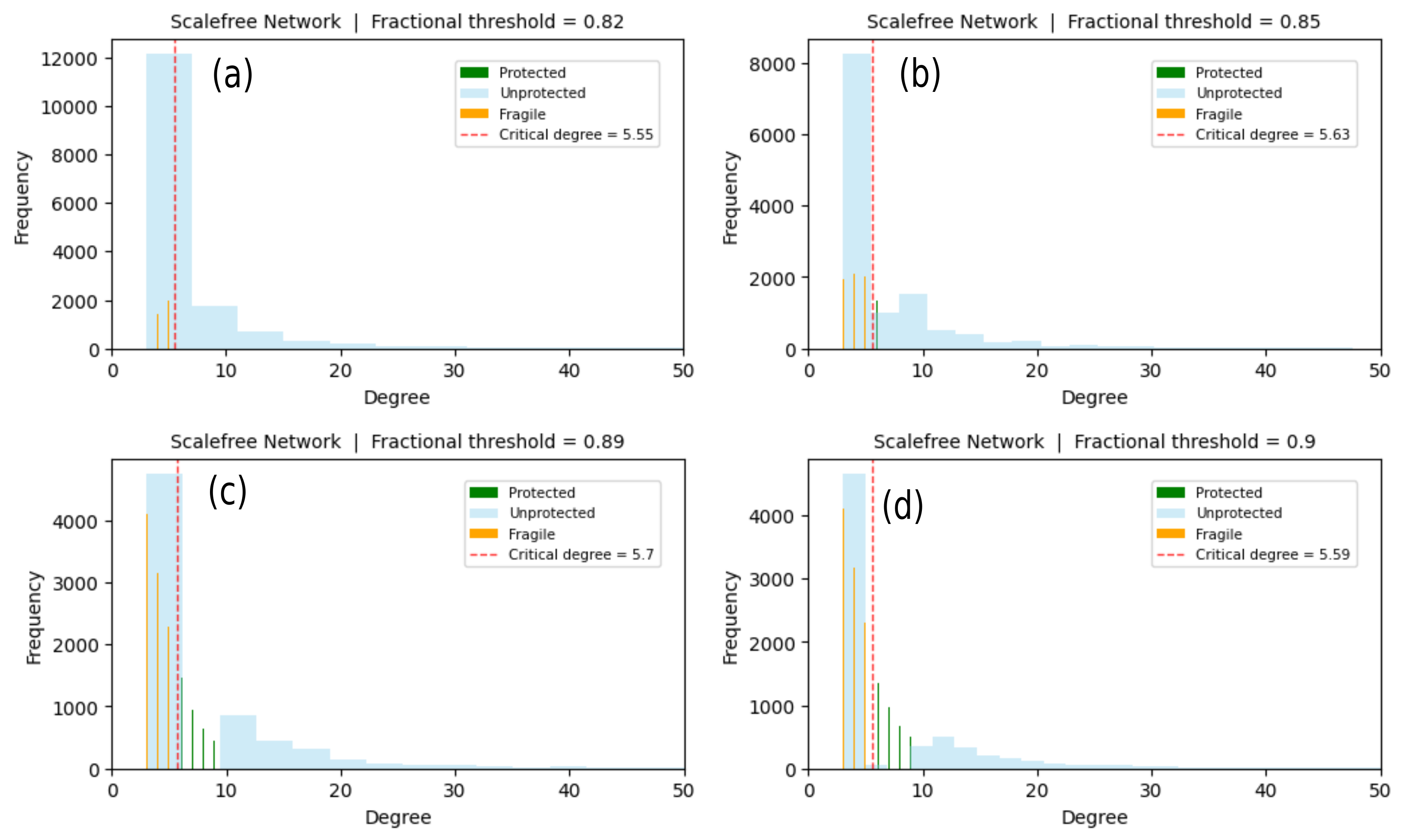}
    \caption{Similar analysis for Scalefree network consisting of $N =20000$ nodes and average degree $<k>=6$.(a) At a fractional threshold of 0.82, no protection is required as all fragile nodes are below the critical threshold (dashed-red line). (b-d) From thresholds 0.835 to 0.89, an increasing number of fragile nodes lie above the critical threshold and therefore need to be protected (green bars).}
    \label{SI_4}
\end{figure}

\section{Critical threshold degree and degree distribution of different types of nodes}
 The critical threshold degree is the cutoff degree above which fragile nodes are designated for protection for effective mitigation. This threshold is determined by coloring the graph~\cite{formanowicz2012survey}, calculating the average degree of each color group, and selecting the lowest average degree as the critical threshold degree. Fragile nodes with a degree greater than this critical value are assigned for protection. Figures \ref{SI_3} and \ref{SI_4} illustrate the distribution of degrees among fragile, protected, and unprotected nodes in ER and Scalefree networks at four different fractional thresholds respectively. Fragile nodes (depicted in yellow) are those that fail under a specific failure mechanism defined in step 3 of our algorithm; meanwhile, protected nodes (displayed in green) refer to fragile nodes with a degree greater than the critical threshold (represented by a red-dashed line), making them eligible for protection. Finally, unprotected nodes are the rest of the nodes that are not fragile(displayed in sky blue).

\end{document}